\documentclass[twocolumn,preprintnumbers,amsmath,amssymb]{revtex4}

\usepackage{epstopdf}
\usepackage{graphicx}
\usepackage{dcolumn}
\usepackage{bm}
\usepackage{amsmath}
\usepackage{hyperref}

\newcommand{\be}{\begin{equation}}
\newcommand{\ee}{\end{equation}}
\newcommand{\beq}{\begin{eqnarray}}
\newcommand{\eeq}{\end{eqnarray}}

\def\barnue{\mathrel{{\bar \nu}_e}}
\def\barnumu{\mathrel{{\bar \nu}_\mu}}
\def\barnutau{\mathrel{{\bar \nu}_\tau}}

\def \lta {\mathrel{\vcenter{\hbox{$<$}\nointerlineskip\hbox{$\sim$}}}}
\def \gta {\mathrel{\vcenter{\hbox{$>$}\nointerlineskip\hbox{$\sim$}}}}

\def\t13{\mathrel{{\theta_{13}}}}
\def\y12{\mathrel{{\tan^2 \theta_{12}}}}
\def\c2{\mathrel{{\chi^2 }}}

\def\ep{\mathrel{{\epsilon}}}

\def\ep{\mathrel{{\epsilon}}}

\newcommand{\df}{DSN$\nu$F}
\newcommand{\snr}{SNR}

\newcommand{\sn}{supernova}
\newcommand{\sne}{supernovae}

\begin{document}

\preprint{INT PUB 06-39}

\title{Testing neutrino spectra formation in collapsing stars with the diffuse supernova neutrino flux}

\author{Cecilia Lunardini}
 \email{lunardi@phys.washington.edu}   
\affiliation{Institute for Nuclear Theory and Department of Physics, University of Washington, Seattle, WA 98195 }%

\date{\today}

\begin{abstract}
  I address the question of what can be learned from the observation
  of the diffuse supernova neutrino flux in the precision phase, at
  next generation detectors of Megaton scale.  An analytical study of
  the spectrum of the diffuse flux shows that, above realistic
  detection thresholds of 10 MeV or higher, the spectrum essentially reflects the
  exponential-times-polynomial structure of the original neutrino
  spectrum at the emission point. There is only a weak (tens of per
  cent) dependence on the power $\beta$ describing the growth of the
  supernova rate with the redshift.  Different original neutrino
  spectra correspond to large differences in the observed spectrum of
  events at a water Cerenkov detector: for typical supernova rates,
  the ratio of the numbers of events in the first and second energy
  bins (of 5 MeV width) varies in the interval 1.5 - 4.3 for pure
  water (energy threshold 18 MeV) and in the range 1 - 2.5 for water
  with Gadolinium (10 MeV threshold).  In the first case
  discrimination would be difficult due to the large errors associated
  with  background. With Gadolinium, instead, the reduction
  of the total error down to $10-20\%$ level would allow spectral
  sensitivity, with a dramatic improvement of precision with respect
  to the SN1987A data. Even in this latter case, for typical neutrino
  luminosity the dependence on $\beta$ is below sensitivity, so that
  it can be safely neglected in data analysis.

\end{abstract}

\pacs{97.60.Bw,14.60.Pq}
\maketitle

\section{Introduction}
\label{intro}

Almost twenty years after the detection of neutrinos from SN1987A,
there is a hunger for new neutrino data from supernovae, needed to
advance our understanding of the physics of core collapse and precious
to improve our knowledge of neutrino physics.

Considering how rare supernovae are in our immediate galactic
neighborhood \cite{Arnaud:2003zr,Ando:2005ka}, it is likely that the
next supernova neutrino data will come to us not in the form of a high
statistics signal from a nearby star, but as a diffuse flux of
neutrinos from all the supernovae in the sky, each of them too far to
give a significant signal.

There are reasons to believe that the detection of this diffuse
supernova neutrino flux (\df) may soon be a reality.  Indeed, the
SuperKamiokande (SK) experiment has already obtained an upper bound on
this flux within an order of magnitude of the theoretical predictions:
1.2 electron antineutrinos ($\barnue$) ${\rm s^{-1} cm^{-2}}$
\cite{Malek:2002ns} (see also
\cite{Eguchi:2003gg,Lunardini:2006sn,Aharmim:2006wq} for other, less
constraining, experimental results).  The sensitivity of SK will
increase substantially in the next 5-10 years with the Gadolinium
addition, GADZOOKS \cite{Beacom:2003nk}.  In this new configuration,
the rate of events due to the \df\ could reach ${\cal O}(1)$
events/year, which implies the possibility to obtain the very first
detection of the diffuse flux. If the supernova rate is precisely
known, GADZOOKS could also give information on the neutrino spectrum
at least for part of the parameter space, with sensitivity comparable
to that of the SN1987A data \cite{Yuksel:2005ae}.

After SK and GADZOOKS, a new generation of detectors with larger
volumes will become operational.  Some projects will require $\sim 1$
Megaton water volumes (HyperKamiokande, UNO, MEMPHYS
\cite{Nakamura:2003hk,Jung:1999jq,Mosca:2005mi}), while other ideas
involve a 50 kiloton mass of liquid scintillator (LENA,
\cite{MarrodanUndagoitia:2006re,Wurm:2007cy}) and 100 kiloton-scale liquid Argon
detectors \cite{Ereditato:2005yx,Cline:2006st}.  For values of the
\df\ that are considered typical, these new detectors are expected to
accumulate a statistics of several tens or even hundreds of events
from the \df\ in the space of few years running time.  

It is clear, thus, that the advent of these new detectors will make it
possible to go beyond the phase of discovery and start a phase of
detailed study of the diffuse flux.  The number of events, their
energy spectrum, and their spatial and temporal distributions will be
analyzed in depth to study  the neutrino
emission from a supernova, the local and cosmological supernova
rate,  neutrino oscillations and  possible exotic particle physics
affecting neutrinos.  

The question of what will be possible to learn from the \df\ data in
the precision phase, and how well, has received only limited
attention.  A initial study by Yuksel, Ando and Beacom
\cite{Yuksel:2005ae} considered specific examples with focus on the
low statistics of GADZOOKS. The main result was that GADZOOKS has a
chance to test the neutrino spectrum if the flux is close to the
current SuperKamiokande limit.  The status of the art on what data
from the \df\ will tell us is well summarized in the work by Fogli et
al.  \cite{Fogli:2004ff}, where a detailed study of the background is
presented for detection with pure water and with water with
Gadolinium.

In this paper I expand on previous works  by discussing in detail what can
be learned from the {\it energy spectrum} of the data from the diffuse
flux, with focus on Megaton scale detectors.  The energy distribution of events depends on the  spectrum
of the neutrinos emitted by an individual supernova and therefore its study would
be a crucial test of models of neutrino transport inside the collapsed
matter of a supernova, as well as of neutrino oscillations inside the
star.  The observed energy spectrum also depends on the distribution
of supernovae with the redshift, i.e. on the dependence of the
supernova rate with the  distance (or, equivalently,
 time from the Big Bang).  Indeed, the larger  the fraction of \sne\ at
cosmological distance, the softer the \df\ spectrum is, due to the
effect of redshift of energy.  To get information on the evolution of the
supernova rate with the cosmological time from the diffuse neutrinos would be of extreme  interest.

Here I present an analytical study of the \df\ spectrum as well as
numerical results for the spectrum of $\barnue$-induced events at a
Megaton water Cerenkov detector. The first offers insight on the
latter.  Of both the \df\ spectrum and the observed signal, I examine
the dependencies on the power of evolution of the supernova rate with
the distance and on the average energy and spectral shape of the
neutrino flux at the source.  I also give typical numbers of events
and discuss the sensitivity of specific observables to the neutrino
spectrum, considering the errors associated with the background and possible uncertainties on the supernova rate normalization.

The paper opens with generalities (Sec. \ref{generalities}), followed
by the analytical study of the \df\ spectrum in sec. \ref{spectrum}.
The observable signal is discussed in Sec. \ref{observables}.  Summary
and discussion close the paper in Sec. \ref{discussion}.

\section{Generalities}
\label{generalities} 

A core collapse supernova is an extremely powerful neutrino source, releasing about $3 \cdot 10^{53}$ ergs of energy within $\sim$10 seconds in neutrinos and antineutrinos of all flavors.  The spectrum of these neutrinos is approximately thermal, with average energies in the range of $10-20$ MeV and with
the muon and tau species being harder than the electron ones due to their weaker (neutral current only) coupling to matter.   Inside the star, the neutrinos undergo either partial or total flavor conversion depending on the mixing angle $\theta_{13}$ and on the mass hierarchy (ordering) of the neutrino mass spectrum (see e.g., \cite{Dighe:1999bi,Lunardini:2003eh}).  For definiteness, here I focus on the $\barnue$ species, as it is the most relevant for detection (see Sec. \ref{observables}).

\begin{figure}[htbp]
\includegraphics[width=0.50\textwidth]{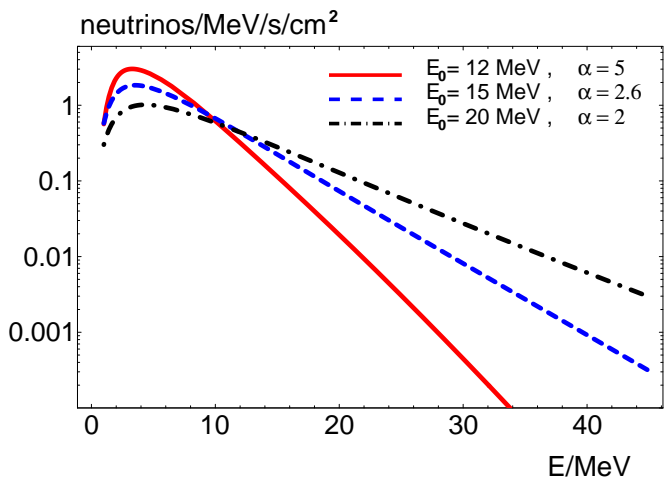}
\caption{Examples of spectra of the diffuse supernova neutrino flux for different original neutrino spectra and fixed supernova rate. For the latter, I used the parameters  $\beta=3.28$, $R_{SN}(0)=10^{-4}~{\rm Mpc^{-3} yr^{-1}}$. The luminosity $L=5 \cdot 10^{52}~{\rm ergs}$ (see Eq. (\ref{nuspec})) was adopted.}
\label{illustration}
\end{figure}

 The energy spectrum of $\barnue$  as they exit the star can be described as \cite{Keil:2002in}:
\be
 \frac{{d} N}{{d} E}\simeq \frac{(1+\alpha)^{1+\alpha}L}
  {\Gamma (1+\alpha){E_{0}}^2}
  \left(\frac{E}{{E_{0}}}\right)^{\alpha}
  e^{-(1+\alpha)E/{E_{0}}},
  \label{nuspec}
\ee where $E_0 \sim 9 - 20$ MeV is the average energy and $L$ the (time-integrated)
luminosity in $\barnue$, $L \sim 5 \cdot 10^{52}$ ergs.  $\alpha$ is a
parameter describing the shape of the spectrum, $\alpha \simeq 2-5$
\cite{Keil:2002in}, with larger $\alpha$ corresponding to narrower spectrum.
The form (\ref{nuspec}) is justified by a number of arguments. First,
it is a good fit to numerical calculations of neutrino transport
before flavor conversion.  Given this, Eq. (\ref{nuspec}) also
describes the $\barnue$ spectrum {\it after} conversion if the
conversion is total, which happens for inverted mass hierarchy and
$\sin^2 \theta_{13}\gta 10^{-4}$ \cite{Dighe:1999bi,Lunardini:2003eh}, or if the $\barnue$ and
$\barnumu,\barnutau$ original spectra are very similar. This is
favored by recent finding that the $\barnue$ and $\barnumu,\barnutau$
average energies may differ only by $10-20\%$ \cite{Keil:2002in}. 
  If the
conversion of $\barnue$ is partial and the hierarchy between the
electron and non-electron original spectra is strong,
Eq. (\ref{nuspec}) is still an accurate description of the neutrino
spectrum above its average energy, where the flux is dominated by the
harder component from oscillated muon and tau neutrinos
\cite{Lunardini:2003eh}.  This condition is met above the realistic
detection thresholds of $\sim 10-20$ MeV energy, since the \df\
average energy is typically smaller than $\sim 5-6$ MeV (see
fig. \ref{illustration}).  

I have estimated the error associated with approximating a composite
spectrum with the form (\ref{nuspec}), in the case of maximal
permutation of the neutrino fluxes, which is realized (for
antineutrinos, of interest here) for the normal mass hierarchy, or for
the inverted hierarchy with $\sin^2 \theta_{13}\lta 3 \cdot 10^{-6}$
\cite{Dighe:1999bi,Lunardini:2003eh}.  I assumed a difference in the
$\barnue$ and $\barnumu,\barnutau$ average energies up to $30\%$ and
up to a factor of two difference in the luminosities of the electron
and non electron flavors.  The result is that the error is smaller
than 6\% above 10 MeV of energy, if Eq. (\ref{nuspec}) is used with
$L$ and $E_0$ being the luminosity and the average energy of the
composite spectrum, and $\alpha$ chosen to reproduce the width of the
composite spectrum.

Finally, the form (\ref{nuspec}) is
supported, as phenomenological description, by the fact that it fits
well the data from SN1987A \footnote{In \cite{Lunardini:2005jf} it was pointed out
that a spectrum of the type (\ref{nuspec}) misses certain spectral
shapes that also fit the SN1987A data well.  Those match
Eq. (\ref{nuspec}), however, above the average energy of the spectrum,
which is the regime of interest here.}.

Given the neutrino output of an individual supernova, one has to sum
over the supernova population of the universe.  This is described by
the cosmological rate of supernovae, $R_{SN}(z)$, defined as the
number of supernovae in the unit time per unit of comoving volume at
redshift $z$.  Its value today ($z=0$) is estimated in the range of
$R_{{ SN}}(0)\sim 10^{-4}~{\rm Mpc^{-3} yr^{-1}}$, with uncertainty of a
factor of two or so \cite{Lunardini:2005jf,Hopkins:2006bw} due to finite statistics and uncertain
systematic errors due to dust extinction.  Theory \cite{Hernquist:2002rg} and data indicate (see e.g. \cite{Hopkins:2006bw})
that after the beginning of star formation the supernova rate (\snr\
from here on) was first somewhat constant in time, and then decreased
after the redshift dropped below $z\sim 1$.  It
can be described phenomenologically with a broken power law:
\begin{eqnarray}
{ R}_{{ SN}}  (z) &=&   { R}_{{ SN}}(0) (1+z)^\beta \hskip 1truecm {\rm for~~z\leq 1} \nonumber\\
&=& { R}_{{ SN}}(0) 2^\beta \hskip 1.9truecm {\rm for~~z>1} ~,
\label{snr}
\end{eqnarray}
where $\beta$ is in the range $\beta \sim 2 - 5 $ from the statistical
analysis of supernova observational data only
\cite{Lunardini:2005jf}. It becomes more constrained around $\beta\sim
3$ (best fit value $\beta=3.28$) if the more precise (but only
indirectly related to the \snr, see
\cite{Ando:2004hc,Lunardini:2005jf}) measurements of the star
formation rate are used instead \cite{Hopkins:2006bw}.  Here I will
consider the wider interval $\beta \sim 2 - 5 $ to be conservative.

Using the supernova rate $R_{{ SN}}$ and the individual neutrino flux, $dN/dE$,  one finds the diffuse flux, differential in energy: 
 \be
\Phi(E)=\frac{c}{H_0}\int_0^{z_{ max}} R_{ SN}(z)\frac{{d}
N(E^\prime)}{{d} E^\prime} \frac{{d}
z}{\sqrt{\Omega_{ m}(1+z)^3+\Omega_\Lambda}} ~,
  \label{flux}
\ee
(see e.g. \cite{Ando:2004hc}).  Here $E$ and
$E^\prime=(1+z) E$ are the neutrino energy at Earth and at the
production point; $\Omega_{ m}\simeq 0.3 $ and $\Omega_\Lambda\simeq
0.7$ are the fraction of the cosmic energy density in matter and dark
energy respectively; $c$ is the speed of light and $H_0$ is the Hubble
constant.  The maximum redshift $z_{max}$ accounts for the expected decrease of the \snr\ with $z$ beyond  $z\sim 5$ or so.  The dependence of the \df\ above realistic detection thresholds on $z_{max}$ is weak due to the effect of redshift.  I will use $z_{max}=5$ from here on. 
 
Throughout the paper, the expression (\ref{flux}) will be used for the
\df.  This formula is valid in the continuum limit, and only if
individual variations between different stars are negligible. This
last statement has not been checked in detail; it is however supported
by the results of numerical simulations \footnote{Basic physics
arguments as well as numerical models indicate that the main features
of a supernova neutrino burst are fairly universal, with only little
dependence on the progenitor mass. However some dependence, at the
level of ten per cent or so, in the average energy and luminosities is
expected in the neutrino emission during the accretion phase, and has
been seen in the output of numerical codes, see
e.g. \cite{Liebendoerfer:2002ny,Kitaura:2005bt}.}.

From Eqs. (\ref{nuspec}), (\ref{snr}), and (\ref{flux}) one sees  that the spectrum of the \df\ depends  on three variables: $\alpha$, $E_0$ and $\beta$, i.e. on the spectrum of the neutrinos emitted by an individual star (``original neutrino spectrum" from here on) and on the power of growth of the \snr\ with $z$.    Fig. \ref{illustration} shows the \df\ spectrum from Eq. (\ref{flux}) for different original neutrino spectra;  in the figure $L$, $R_{SN}(0)$ and $\beta$ are held fixed while $\alpha$ and $E_0$ are varied (see caption for details) to show how the slope of the spectrum can change.  Specifically, the spectrum is steeper for smaller $E_0$ and larger $\alpha$, as it will be discussed in detail  in Sec. \ref{spectrum}.

 \section{The energy spectrum of the diffuse flux}
 \label{spectrum}

What is the energy distribution of the diffuse supernova neutrinos? How
is it related to the original neutrino spectrum and to the cosmological distribution of supernovae?  In the
light of what discussed in Sec. \ref{generalities}, the answer is only
partially intuitive. 
 Due to the redshift of energy, the farther a \sn\ is from us the more
its contribution is shifted to lower energy. This means that the \df\
does not retain the same spectrum that the neutrinos had at emission;
but have instead a distorted -- and overall softer -- distribution
with respect to it, which depends on the \snr.  One expects the
largest distortion in the lowest energy part of the spectrum $E \sim
5-10$ MeV, where the contribution of high redshift supernovae is
larger. In contrast, in the high energy end of the \df\ spectrum, $E
\gta 15-20$ MeV, the flux is largely dominated by stars with
negligibly small redshift, and therefore it should be closer to the
original spectrum.
 
Let us see this in a more quantitative way, by examining  the spectrum that results from Eq. (\ref{flux}).   There, the integration can not be done exactly. Still, however,  one can find useful analytical approximations that help to understand the energy dependence of the \df. 
Considering that current and planned searches of the diffuse flux have thresholds of $\sim 10$ MeV or higher, it is sensible to focus on the high end of the spectrum, i.e. on the small redshift approximation, $z \ll 1$. In this limit one can make the following simplifications: 
\begin{enumerate}

\item The substitution:
\be
\left[\Omega_{ m}(1+z)^3+\Omega_\Lambda \right]^{-1/2} \rightarrow (1+z)^{-\frac{3}{2}\Omega_m}~.
\label{furbata}
\ee
The two expressions coincide at first order in $z$ under the condition that $\Omega_m + \Omega_\Lambda=1$, which is favored by current cosmological data. 

\item Neglect  the contribution  of the \sne\ beyond the $z \simeq 1$ break of the \snr, specifically:
\begin{eqnarray}
{ R}_{{ SN}}  (z) &=&   { R}_{{ SN}}(0) (1+z)^\beta \hskip 0.6 truecm {\rm for~~z\lta z_{max}}~, \nonumber\\
&=& 0 \hskip 2.85truecm {\rm for~~z>z_{max}} ~,
\label{eq:powerlaw}
\end{eqnarray}
with $z_{max} = 1$.
\end{enumerate}

With these approximations, the integral in Eq. (\ref{flux}) gives the following result:
\begin{widetext}
\be
\Phi_e \simeq  { R}_{{ SN}}(0) \frac{c}{H_0} \frac{ L}
  {\Gamma (2+\alpha){\epsilon}^2}
  \left(\frac{E}{\epsilon} \right)^{\alpha-\eta-1} \left( \Gamma \left[ \eta+1, {E}/{\ep} \right] - \Gamma\left[ \eta+1, (1+z_{max}){E}/{\ep} \right] \right)~,
\label{generalresult}
\ee
\end{widetext}

where I define $\eta \equiv \alpha+ \beta - 3 \Omega_m/2$ and   $\epsilon=E_0/(1+\alpha)$ for brevity. Considering $\alpha \gta 2$ (Sec. \ref{generalities}), we have $\epsilon \lta E_0/3$.

It is possible to further simplify the result by dropping the
contribution of the upper limit of integration $(1+z_{max})
E/\ep$, which at high energy is exponentially suppressed with respect
to the term due to the lower limit.  Furthermore, if $\eta$ is integer, one can also use the expression of the $\Gamma$ function in terms of an exponential times a polynomial~\footnote{I recall the expressions of the function $\Gamma$:
$ \int dx ~x^w e^{-x} = -\Gamma(w+1,x)~, ~ \Gamma(n+1,x)=n ! e^{-x} \sum^n_{k=0} \frac{x^k}{k !}~$,
the second being valid only for integer $n$.}, and get:
\begin{widetext}
\beq
\Phi_e& \simeq & { R}_{{ SN}}(0) \frac{c}{H_0} \frac{ L}
  {\Gamma (2+\alpha){\epsilon}^2}
~e^{-\frac{E}{\epsilon}} \sum^\eta_{k=0} \left[ \left(\frac{E}{\epsilon} \right)^{\alpha -1-k} \frac{\eta!}{(\eta-k)!}   \right]~,
\label{crudecompact}
\eeq
or,  
explicitly:
\beq
\Phi_e& \simeq & { R}_{{ SN}}(0) \frac{c}{H_0} \frac{ L}
  {\Gamma (2+\alpha){\epsilon}^2}
~e^{-\frac{E}{\epsilon}} \nonumber\\
&& \left[  \left(\frac{E}{\epsilon} \right)^{\alpha-1} + \eta \left(\frac{E}{\epsilon} \right)^{\alpha -2} +  \eta (\eta -1) \left(\frac{E}{\epsilon} \right)^{\alpha -3} +  \eta (\eta -1)(\eta -2) \left(\frac{E}{\epsilon} \right)^{\alpha -4} + ... \right]~.
\label{crude}
\eeq
\end{widetext}
Eq.  (\ref{crude}) turns out to be numerically accurate 
also for non-integer $\eta$, provided that the sum is truncated at the integer nearest to $\eta$. It also has the advantage of simplicity, so it can be used to understand the physics.  

Let us pause for a moment and study the result (\ref{crude}).  This
expression tells us that the \df\ spectrum is similar to the original spectrum in the general  structure of an exponential times a polynomial part.   More specifically, the term $e^{-E/\epsilon}/\epsilon^2$ in Eq. (\ref{crude}) is the same as in the original spectrum, while the polynomial part is more complicated and contains the information on the cosmology (i.e., on $\beta$ and on $\Omega_m$), through the quantity $\eta$.     The latter enters the result only in the
coefficients of the various powers of energy, a very weak dependence
with respect to the exponential decay of the spectrum in $E/\ep$. 
Notice that, in agreement with intuition, the dependence on $\eta$ becomes weaker for higher energy: indeed, in the limit $E \gg \epsilon$, the polynomial part is dominated by the $\eta$-independent term $(E/\epsilon)^{\alpha-1}$.
 In
general, Eq. (\ref{crude}) suggest that the exponential captures the
most of the energy dependence, with only slower variations due to the
polynomial part.     Thus, one can consider fitting the \df\ spectrum at high energy with a simple exponential form, as already suggested in \cite{Malek:2003ki}:
\be 
\Phi_e=\Phi_e(0) e^{-E/\langle E \rangle}~.
\label{crudest}
\ee
I find that, expectedly, $\langle E \rangle$ is numerically close to $\ep$, with differences smaller than $\sim 30\%$.

\begin{figure}[htbp]
  \centering
   \includegraphics[width=0.50\textwidth]{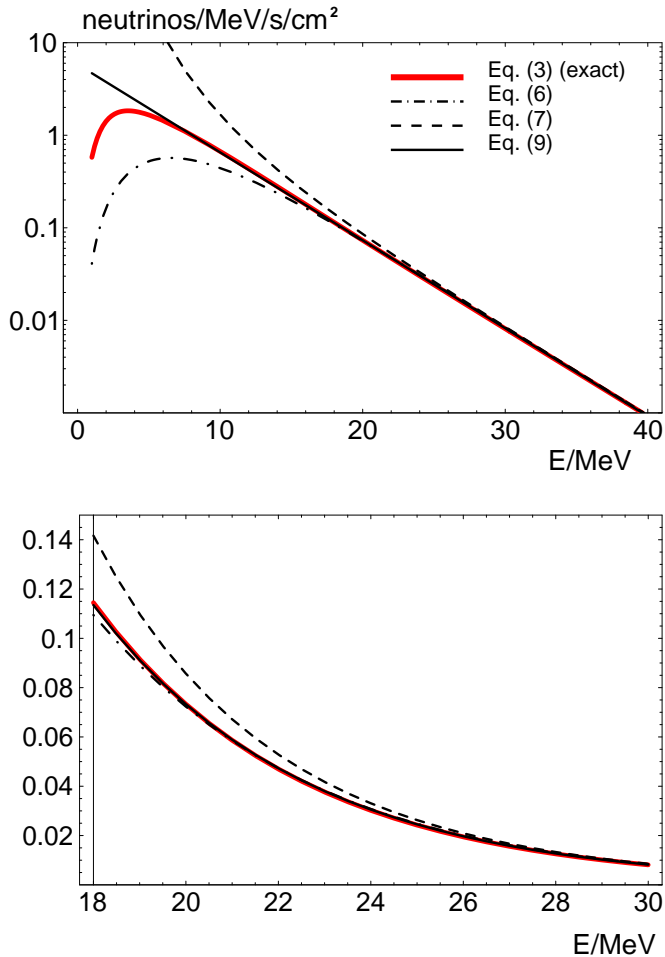}
\caption{An illustration of the approximate analytical descriptions of the \df\ spectrum in comparison with the exact spectrum calculated numerically (details in the legend).  The lower panel is a close-up of the upper one on linear scale.
I used the same \snr\ parameters and luminosity as in fig. \ref{illustration} and the spectral parameters $E_0= 15$ MeV, $\alpha=2.6$.   For the fitted exponential form, Eq. (\ref{crudest}), the value of the characteristic energy is $\langle E \rangle=4.57$ MeV.}
\label{approximations}
\end{figure}

Fig. \ref{approximations} compares the exact spectrum (see caption for
details), calculated numerically, with the three approximations: the
sophisticated one, Eq. (\ref{generalresult}), the cruder one,
Eq. (\ref{crudecompact}), and the phenomenological one,
(\ref{crudest}). For the latter, the parameters $\langle E \rangle$
and $\Phi_e(0)$ have been adjusted to fit the numerical curve in the
interval $E \simeq 20-30 $ MeV.  The figure shows that the
sophisticated result, Eq. (\ref{generalresult}), underestimates the
flux at low energy as a result of neglecting $z>1$ part of the
\snr. Instead, the the cruder description, Eq. (\ref{crudecompact}),
overestimates the flux at low energy due to dropping the (negative)
contribution of the upper limit of integration in
(\ref{generalresult}), which is equivalent to taking the \snr\ as growing with constant power all the way to $z\rightarrow \infty$. The simple fitted exponential form lies between the two approximations and works surprisingly
well down to $E\sim 5-6$ MeV, probably because by fitting $\langle E \rangle$ one
effectively reproduces at least part of the contributions that are not included in the forms (\ref{generalresult}) and (\ref{crudecompact}), especially the $z>1$ part of the \snr.

I have studied the precision of Eqs. (\ref{generalresult}) and
(\ref{crudecompact}) in detail.  First, I fixed $\beta=3.28$, and
varied the neutrino spectral parameters in the intervals $\alpha = 2
-5 $ and $E_0 = 9 -18$ MeV.  I obtained that the full result,
Eqs. (\ref{generalresult}), is precise to better than $12\%$ (40\%) at
$E=19.3$ MeV ($E=11.3$ MeV).  Instead, the cruder expression
(\ref{crudecompact}) exceeds the exact result by up to 40\% at
$E=19.3$ and up to a factor of three at $E=11.3$ MeV.
 
The analytical results lose accuracy for larger $\beta$, as expected, since larger $\beta$ means larger contribution of supernovae at $z>1$.   With $\beta=5$, the sophisticated result (\ref{generalresult}) errs by up to $20\%$ (120\%) at
$E=19.3$ MeV ($E=11.3$ MeV). Eq. (\ref{crudecompact}) can deviate from the exact result by a factor of several.

The exponential form, Eq. (\ref{crudest}), with $\langle E \rangle$ fitted to the reproduce the  exact result at $E\sim 20-30$ MeV, is accurate within $5\%$ everywhere above 20 MeV and deviates from the exact spectrum by at most $25\%$ at $E=11.3$ MeV. 
 
In closing this section, I would like to emphasize that, while useful
for understanding, the analytical approximations discussed here lose
validity below 15-20 MeV of energy.  Thus, to model observables at
these low energies, it is necessary to calculate the neutrino flux
numerically using Eq. (\ref{nuspec}).  I do so in the reminder of this
work.

\section{Detecting the diffuse flux: observables}
\label{observables}

\subsection{Inverse beta decay: signal and background}
\label{ibd}

I now turn to discussing the spectrum of events induced by the diffuse
flux in a detector. In particular, I consider events due to inverse
beta decay, $\barnue + p \rightarrow n + e^+$, since they largely
dominate the signal in water and liquid scintillator. 

The detected positrons from
inverse beta decay carry information on the incoming $\barnue$
spectrum. This is thanks to the fact that, except for  small effects due to
nucleon recoil, the neutrino and positron energy differ
simply by the mass difference of neutron and proton:
$E_{e^+}=E-1.29~{\rm MeV}$.  Thus, the positron energy spectrum is
roughly given by the neutrino spectrum distorted by the energy
dependence of the detection cross section, which goes approximately as
$E^2$. 

 To calculate
the spectrum of the observed positrons, I model the diffuse $\barnue$
flux according to Eq. (\ref{nuspec}), with $z_{max}=5$. In all cases,
I will use the typical values $R_{SN}(0)=10^{-4}~{\rm Mpc^{-3}
yr^{-1}}$ and $L=5 \cdot 10^{52}~{\rm ergs}$, and vary $\alpha$, $E_0$
and $\beta$ in the intervals that I indicated as typical in
Sec. \ref{generalities}.  I use the detection cross section from
ref. \cite{Strumia:2003zx}.

For definiteness, results are given for a water Cerenkov detector of
0.45 Mt fiducial volume (20 times the volume of SK) without and with
Gadolinium, after 4 years of operation.  I consider two different
energy thresholds, 18 MeV and 10 MeV in positron energy, characteristic of water only and water with
Gadolinium (water+Gd hereafter) respectively.  I adopt the same energy
resolution and efficiency 
($\simeq$93\% in the energy range relevant here \footnote{In the SK
  analysis \cite{Malek:2003ki} several experimental cuts reduced the
  efficiency down to 47\% (79\%) below (above) 34 MeV. This could
  change in the future depending on the details of the data
  processing. Thus, for the sake of generality here I consider the
  generic instrumental efficiency of the SK detector.})  as SK \cite{Hirata:1988ad,Bahcall:1996ha}.  The
set ups considered here are the same as those in ref.
\cite{Fogli:2004ff} (except for a minor difference in the volume:
$0.4$ Mt in ref. \cite{Fogli:2004ff} and 0.45 Mt in this work), so
that a direct comparison is possible.

For a meaningful interpretation of the results, it is necessary to
take into account the background carefully.  As described in a number
of publications (e.g., \cite{Ando:2004hc} and references therein), there are four sources of background
that in water result truly ineliminable. These are: spallation
neutrons, reactor neutrinos, atmospheric neutrinos and invisible
muons.  The first one is dealt with by setting the energy threshold at
18 MeV of positron energy.  Above this cut the spallation and reactor
backgrounds are practically zero. The remaining backgrounds,
atmospheric neutrinos and invisible muons, can only be included in the
statistical analysis, as done in  \cite{Malek:2002ns}, so that the
signal and these backgrounds can be separated only on statistical
basis. Of these, the invisible muons background is larger than the \df\ signal by a factor of several \cite{Malek:2002ns,Fogli:2004ff}.

 The addition of Gadolinium to the water will allow to
discriminate inverse beta decay events from events of other type,
thanks to the detection in coincidence of the positron and of the
neutron capture on Gd \cite{Beacom:2003nk}.
With Gd the energy threshold can be lowered to 10 MeV of positron
energy, motivated by the presence of the reactor $\barnue$'s below it.
It is estimated that in the water+Gd configuration the invisible muon
background will be reduced by a factor of $\sim$5 with respect to pure water
 \cite{Yuksel:2005ae}. This reduction will bring it down to be smaller or comparable to the signal above the 10 MeV threshold.

Here I model the background following ref. \cite{Fogli:2004ff}.  As done there,  the events due to background will not be shown, but will be included in the calculation of the statistical error.  More explicitly, I consider a scenario in which from the total of the data one subtracts the background using a Monte Carlo prediction for it. The error on the subtracted signal is then given by the statistical error on the total (signal plus background) number of events:  $\sigma = \sqrt{N_{sig}+N_{bckg}}$.

The results of this work can be easily rescaled to reproduce the
inverse beta decay signal at other detectors (up to differences due to
different energy resolution).  The case of the proposed 50 kt liquid
scintillator detector LENA \cite{MarrodanUndagoitia:2006re,Wurm:2007cy} is particularly interesting.
Since scintillator  distinguishes inverse beta decay through the
detection of the positron and neutron in coincidence, its performance
will be similar to that of a water+Gd detector, up to a rescaling to
account for the different volume.  I estimate that, in terms of number of events, 1 (4) year(s) of running
time of a 0.45 Mt water+Gd detector with the efficiency given above corresponds to $\sim$7.4 ($\sim$30) years running
time of LENA with 100\% efficiency.  Equivalently, for the same running time, LENA will have a number of events smaller by a factor $\sim 0.13$.

\subsection{The positron energy spectrum}
\label{posspec}

\begin{figure}[htbp]
  \centering
 \includegraphics[width=0.50\textwidth]{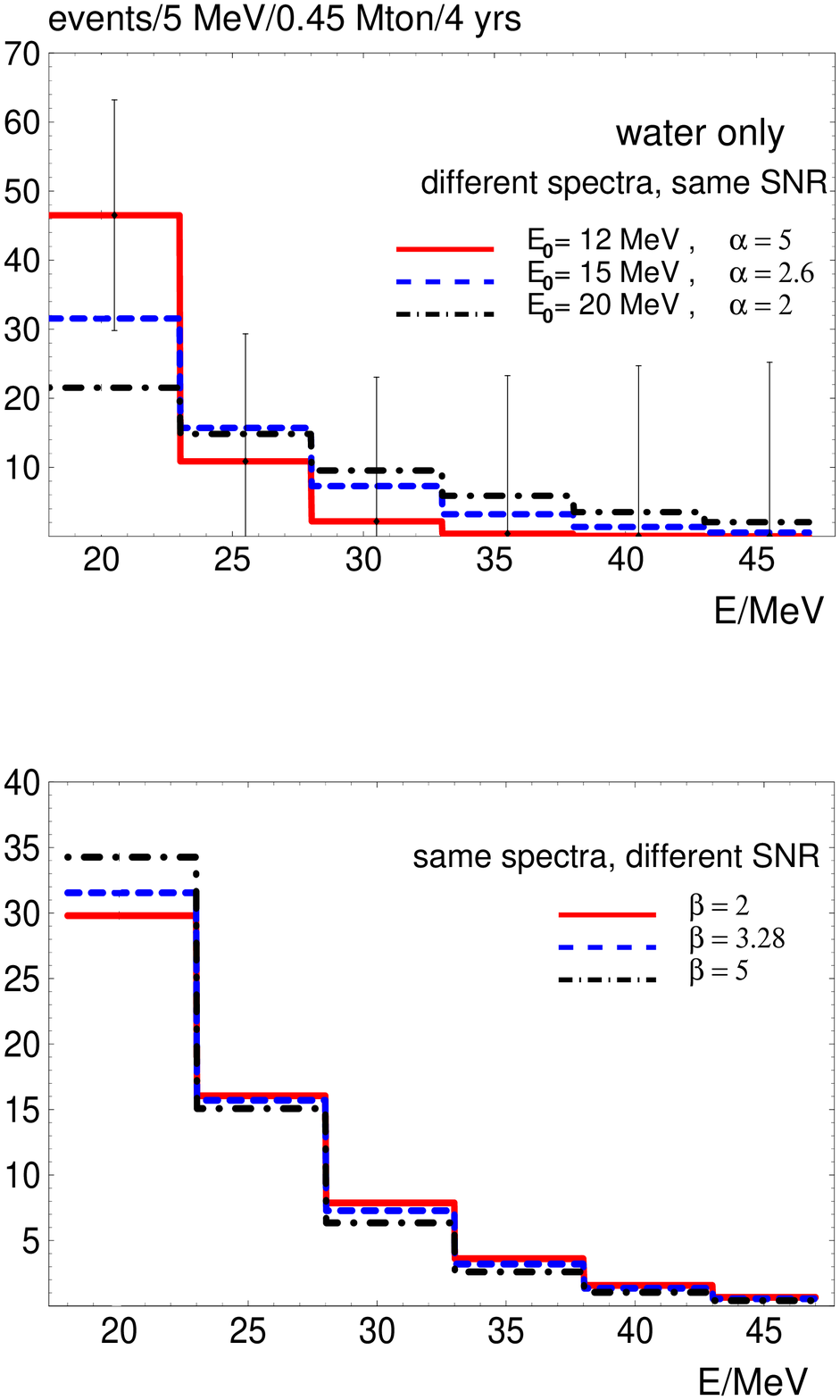}
\caption{The spectra of the observed positrons from inverse beta decay  for the pure water setup, for fixed SNR (upper panel, $\beta=3.28$ used) and for fixed original neutrino spectrum (lower panel, $E_0= 15$ MeV, $\alpha=2.6$ used).  In all cases the numbers of events have been normalized to $N=60$.   }
\label{sk}
\end{figure}

\begin{figure}[htbp]
  \centering
 \includegraphics[width=0.50\textwidth]{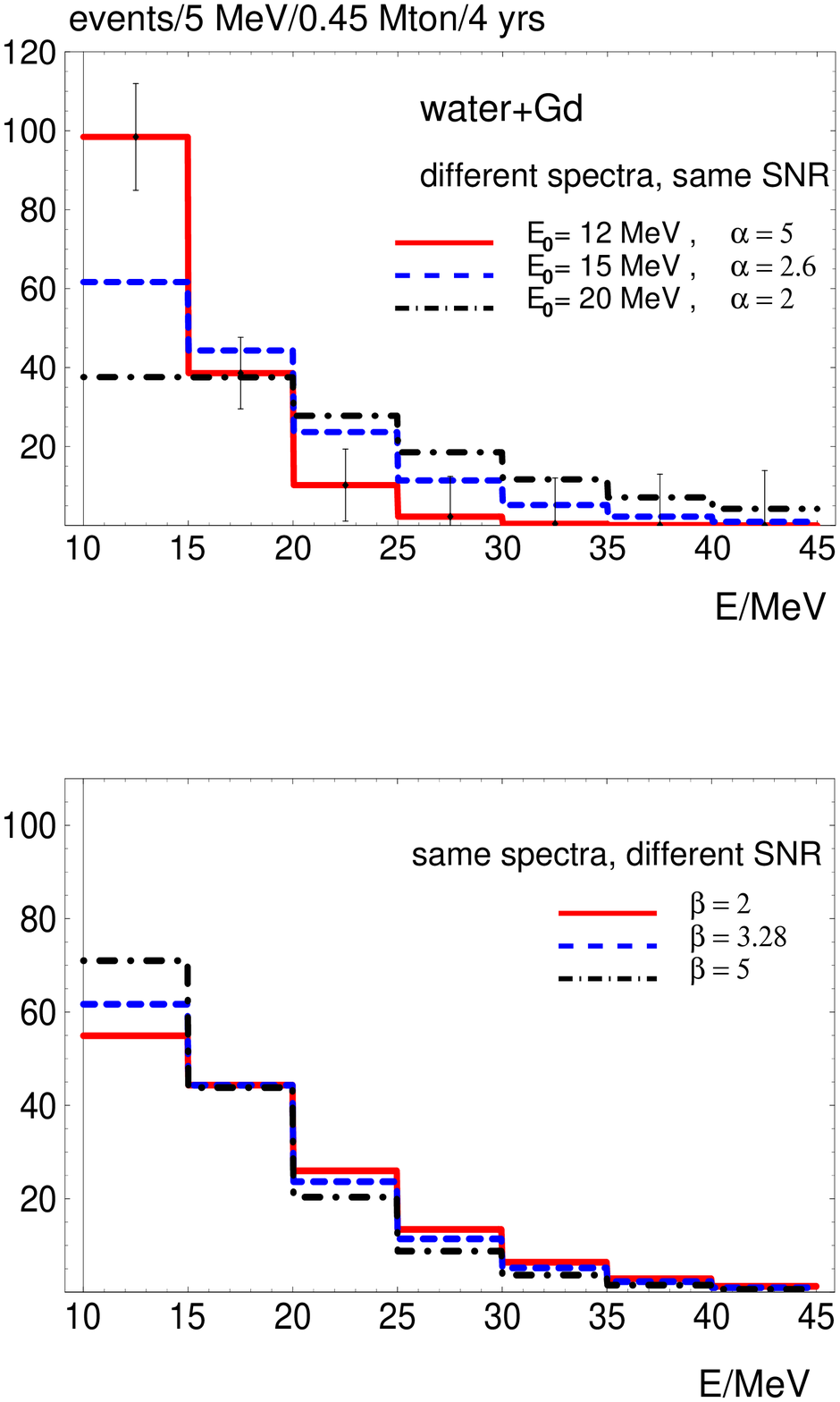}
  \caption{ Same as fig. \ref{sk} for water+Gd. The numbers of events have been normalized to $N=150$ in all cases.}
\label{gadz}
\end{figure}

Figures \ref{sk} and \ref{gadz} give the spectra of events in 5 MeV
bins for the water only configuration and the water+Gd one.  For
visual convenience, in each figure all the histograms have been
normalized at the same number of events, N=60 for water only and N=150
for water+Gd. These are the numbers obtained with $\alpha=2.6$,
$E_0=15$ MeV and $\beta=3.28$.
Each figure has two
panels: in the upper one I fix the cosmology and vary the neutrino
spectral parameters, $\alpha$ and $E_0$, while in the lower panel the
spectral parameters are fixed and $\beta$ is varied (see figure
captions for details). The error bars (omitted in the lower panels for
simplicity) represent the 1 $\sigma$ {\it total} (signal+background)
statistical errors.

Let us first comment on the water only case, Fig. \ref{sk}.  It
appears immediately that the spectrum changes substantially when the
spectral parameters are varied. The behavior agrees with what expected
from Sec. \ref{spectrum}: the spectrum falls more rapidly for the
smaller $\ep$, according to the exponential dependence in the \df\
spectrum.  So, a first conclusion is that in principle the signal
induced by the diffuse flux is very sensitive to the original neutrino
spectrum.

In spite of this,
however, different \df\ spectra -- and therefore different original
spectra -- will be difficult to distinguish one from the other with
high statistical confidence. This appears from the error bars plotted
in fig.  \ref{sk}. One can see that the largest variation of the
numbers of events in the first bin is within 2 $\sigma$ statistical
error, and the variations are smaller in the other energy bins.  As
already anticipated in \cite{Fogli:2004ff}, this means that with typical
statistics a Megaton water detector would be able to give indication
or evidence of the \df, but not to give precise spectral
information about it.  With higher \df\ flux -- within what is allowed
by the current SK limit -- the spectral sensitivity could be
moderately better. For example, with twice as many signal events, the
variation of signal between the two extreme cases in the first energy
bin would be of about $\sim 3\sigma$.  Notice that the total
statistical error is dominated by the invisible muon background and
therefore would increase only weakly with the increase of the signal.

The lower panel of Fig. \ref{sk} confirms the expectation of little
dependence on the \snr\ parameter $\beta$.  Even with the rather
extreme variations of $\beta$ used in the figure, the number of events
varies by at most $\sim 15\%$ in the bins where the statistics is
higher.  Considering the large  errors, I conclude that
very little or no information on $\beta$ could be extracted from \df\
data.

With water+Gd the potential to study the \df\ spectrum is dramatically
better: indeed, the statistical errors are smaller thanks to the
improved background reduction, and the signal is enhanced by the lower
energy threshold.  Fig. \ref{gadz} illustrates this. It appears that
the two extreme cases differ by more than $ 3\sigma$ in the first
bin. This means that the possibility of spectrum discrimination is
concrete, even though the differences in the other bins
are not as significant.
As in the case of pure water, the dependence of the observed spectrum on $\beta$ is weak.  The difference between the
number of events in the first bin for the two extreme scenarios is
at the level of $ \sim 1 \sigma$, increasing to  $\sim (2 -2.5) \sigma$ for doubled or tripled signal.  Thus, to discriminate between different values of $\beta$ would be difficult, even in the ideal case of perfectly known original spectrum. 

The points made above can be illustrated more quantitatively using the ratio $r$ of the number of events in the first and second energy bin.
Taking bins of 5 MeV width  this ratio is:
\be
r \equiv \frac{N(18\leq E_{e^+}/{\rm MeV}<23)}{N(23\leq E_{e^+}/{\rm MeV}<28 )}~,
\label{ratio}
\ee
for the water only case, and 
\be
r \equiv \frac{N(10\leq E_{e^+}/{\rm MeV}<15)}{N(15\leq E_{e^+}/{\rm MeV}<20)}~,
\label{ratio_gd}
\ee for water+Gd, in terms of the positron energy $E_{e^+}$.  To limit
oneself to the numbers of events in the first two bins is overall more
convenient with respect to including the data in the higher energy bins.
The latter add little to the signal but add substantially to the
background, and therefore they end up washing out the effect of the signal by increasing the error.

For pure water and fixed \snr\ (Fig. \ref{sk}, upper panel) I find
that $r$ varies in the range $r \simeq 1.5 - 4.3 $. With the
statistics used in Fig. \ref{sk}, the error on $r$ is larger than
$\sim 100\%$; this confirms the conclusion that the sensitivity to the
neutrino spectrum is limited.  For fixed original spectrum and varying
$\beta$, the variation of $r$ is minimal, much smaller than the error.

For the water+Gd configuration in the case of fixed \snr\ (Fig.
\ref{gadz}, upper panel) I have $r \simeq 1 - 2.5 $. The error is
$\sim 20 - 30 \%$, meaning that at least the extreme cases should be
distinguishable for typical luminosity of the \df. For fixed original
spectrum and varying $\beta$ I get $r \simeq 1.2 - 1.6 $, which is
smaller than the error unless the \df\ is close to the SK upper limit.

\subsection{Number of events}

To complement the study of the \df\ spectrum -- which is the main focus of this paper -- here I give a brief discussion of the event rates and their information content. 

As it is clear from the previous sections, the number of events in a
detector depends on the original neutrino spectrum, on  the \snr\ power
$\beta$ and on the overall factors $R_{SN}(0)$ and $L$, i.e. the
normalization of the \snr\ and the neutrino luminosity in $\barnue$
(after oscillations).  Once the detection of the \df\ moves from the
discovery to the precision phase, the event rate will constrain the product $R_{SN}(0)\times L$ and will complement the spectral analysis to extract $E_0$, $\alpha$ and $\beta$.
Realistically, these two aspects can not be separated: the data will
be analyzed by a global fit with both the spectral parameters and the
overall factors as fit variables.  Still, for illustration it is
useful to discuss certain limiting cases where the sensitivity to the
overall factors and that to the spectral parameters can be treated
separately.

If the \df\ spectrum is reconstructed very precisely by the spectral
analysis, the event rate will serve mainly to constrain
$R_{SN}(0)\times L$.  This can happen for the water+Gd setup, where
the statistical errors are small enough to allow spectral
reconstruction. High precision would be obtained if the \df\ signal is
particularly luminous, typically two or more times larger than in
Fig. \ref{gadz}.  In the ideal case of perfectly known spectrum, the
error on the $R_{SN}(0)\times L$ inferred from the data would be just
the statistical error, of the order of tens of per cent. This would be
an enormous improvement with respect to the current information from
supernova observations (which constrain $R_{SN}(0)$) and from the
neutrino data from SN1987A (see e.g. \cite{Lunardini:2005jf}).  

It should be considered that by the time the \df\ is detected, the
\snr\ could be better known from new supernova surveys like JWST
\cite{jwst} and SNAP \cite{snap}, since these are expected to see
thousands of core collapse supernovae while undertaking their primary
mission of observing type Ia supernovae.  From these surveys we are
likely to obtain a precise measurement of $\beta$, while some
systematic uncertainty on $R_{SN}(0)$ will be left due to extinction
effects. In this situation, the measured $\beta$ will be used as input
in the \df\ spectral analysis, thus increasing the precision of
reconstruction of $E_0$ and $\alpha$.  The astrophysical measurement
of $R_{SN}(0)$ and the information on $R_{SN}(0)\times L$ from the
neutrino event rate will be combined to constrain $L$ and reduce the
systematic error on $R_{SN}(0)$.  If the systematics due to extinction
is well understood and taken into account, the \df\ event rate will
essentially measure $L$.

It is possible that the product $R_{SN}(0)\times L$ and the power
$\beta$ will be much better known than the original neutrino spectrum.
This scenario could be realized if $R_{SN}(0)$ and $\beta$ are
precisely measured by supernova surveys (plus precise treatment of
systematic effects) and $L$ is determined  by
improved numerical simulations of core collapse, perhaps combined with  astrophysical data on supernovae such as energy of the explosion, mass of the remnant, etc..

\begin{figure}[htbp]
  \centering
\includegraphics[width=0.50\textwidth]{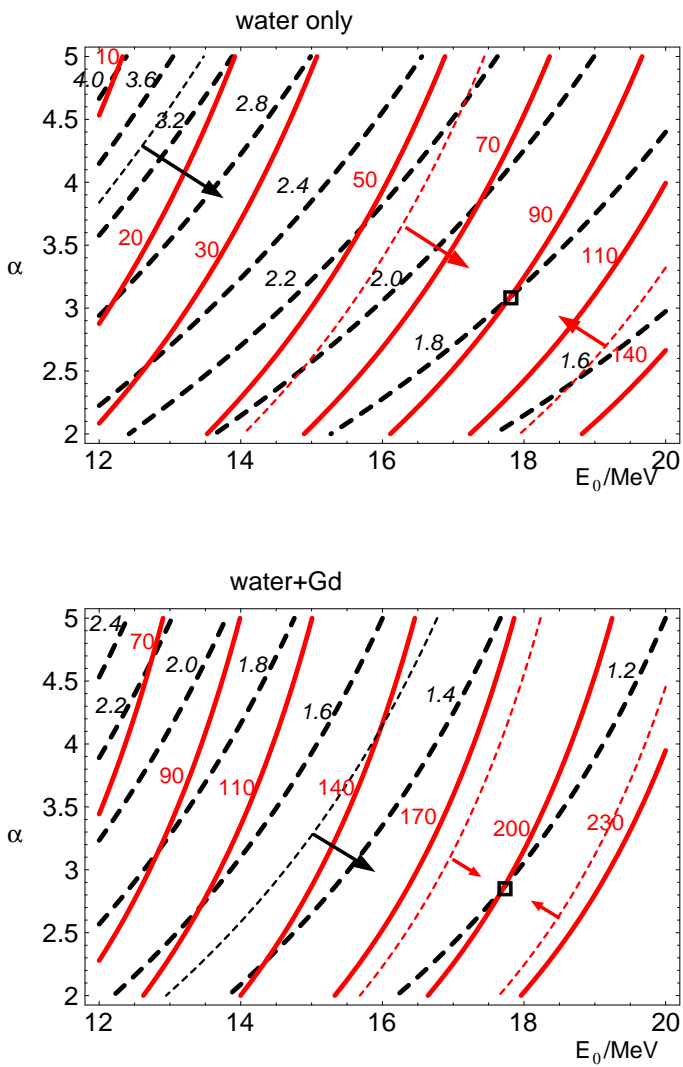}
\caption{Solid (red) lines: isocontours of the number of events in 4 years running time, $N$, in the space $\alpha - E_0$.  Long dashed (black) lines: isocontours of the ratio $r$ of the numbers events in the first and second bin (Eqs. (\ref{ratio}) and (\ref{ratio_gd})). In each panel I show a possible measurement of $N$ and $r$, with central values (diamond) and 1$\sigma$ statistical errors (regions within the short dashed lines, marked with arrows; the wider region is the error on $r$). 
The \snr\ parameters and luminosity are as in fig. \ref{illustration}.}
\label{curvessensitivity}
\end{figure}

In this case the event rate will serve to constrain the
spectral parameters in combination with the energy distribution of the
data. 
%
In fig. \ref{curvessensitivity} I briefly illustrate this scenario in
the ideal situation in which $L$, $R_{SN}(0)$ and $\beta$ are known with
negligible error (see caption for details).  The figure shows
isocontours of the ratio $r$ and of the event rate $N$ in the plane
$\alpha- E_0$, both for water only and for water with Gadolinium. It appears that the  event rate varies by a
factor of several depending on the spectral parameters: from $\sim 10$
to $\sim 140$ ($\sim 70 - 230$) events in 4 years for water only (water plus Gadolinium)
\footnote{These event rates are well compatible with the current upper
limit from SuperKamiokande \cite{Malek:2002ns}.  Such limit allows up
to 3.5 events/year at SK \cite{Malek:2003ki}, corresponding to about
250 events in 4 years in a 0.45 Mt detector in the water only
configuration.}.  The figure also illustrates a possible measurement
of $N$ and $r$, with central values and 1$\sigma$ statistical errors
on these two quantities \footnote{For the number of events I have used
the sum of the numbers of events in the first three (four) energy bins
for pure water (water+Gd). The higher energy bins are
background-dominated and thus including them would increase the error
on the signal.  }.  If only $r$ is considered -- for robustness
against uncertainties on $R_{SN}(0)\times L$ -- then for water only a
major portion of the parameter space is allowed within $1\sigma$, and
the entire space is within $3\sigma$. Instead, for water+Gd the
$1\sigma$ allowed region is much more restricted, and the $3\sigma$
region excludes the edges of the parameter space; for example, the
measurement of $r=1.2$ would exclude the point $E_0=12$ MeV and
$\alpha=4.5$.  Including the information on the number of events
reduces the allowed region substantially: while for pure water the
whole $\alpha- E_0$ plane is allowed within $3\sigma$, for water+Gd
the potential for exclusion is better. For the specific case in the
figure, a measurement of $N=200$ would exclude $E_0=12$ MeV regardless
of the value of $\alpha$.

Notice that the region allowed by $N$ is contained inside the one
allowed by $r$, meaning that the number of events alone is sufficient
to constrain the original neutrino spectrum, if $L$, $R_{SN}(0)$ and
$\beta$ are known very precisely.  In presence of errors on these
quantities, which is probably more realistic, the ratio $r$ will be
more constraining instead.


\section{Summary and discussion} 
\label{discussion}

It is a concrete possibility that the next supernova neutrino data
will be from the diffuse flux.  If so, these data will be of interest to
improve on our knowledge of the supernova neutrino spectrum with
respect to what is already known from SN1987A.  Furthermore, the \df\
data will be analyzed to test the cosmological rate of supernovae.

Here I have addressed the question of the sensitivity of the \df\ to
these quantities, both at the level of the $\barnue$ flux at earth and
of the spectrum of inverse beta decay events in a detector.

One first answer is that the spectrum of the diffuse $\barnue$ at
Earth above realistic detection thresholds (10 MeV or higher)
essentially reflects the neutrino spectrum at the source (after
oscillations), with only a weak (tens of per cent) dependence on the power $\beta$ of evolution
of the supernova rate with the redshift. Indeed, with respect to the
original spectrum, the \df\ spectrum retains the same structure of an
exponential times a polynomial part. The exponential term is
identical to the one in the original spectrum, and dominates the
energy dependence of the \df\ spectrum.  The information on the
cosmology is in the coefficients of the polynomial, and therefore the cosmological parameters
influence the spectrum more weakly.  These features, that were studied
in detail here, are in agreement with the general intuition about the
contribution of more distant supernovae being weak at high energy due
to redshift.

At the level of experiment, a better background subtraction with respect
to pure water (e.g, by the addition of Gadolinium to the water of
UNO/HyperKamiokande/MEMPHYS or by using a large scintillator detector (LENA))
would be needed to move from the discovery phase to the precision
phase.  

What major advances will come from the precision phase depends on the
situation in the field at that time.  A likely scenario is that  there will be new,
precise measurements of the supernova rate. These will result in a
precise determination of $\beta$, while leaving some uncertainty on
the overall scale of the \snr, $R_{SN}(0)$, due to effects of
extinction.  In this case, the \df\ data will mainly improve our
knowledge of the original neutrino spectrum, thus adding precious
information to what is already known from SN1987A. For typical
luminosity of the signal, after 4 years of operation a 0.45 Mt
detector in the water+Gd configuration would have hundreds of signal events,
and total (signal+background) statistical error of $\sim 10-30\%$ in
the first two energy bins. This should make it possible to
discriminate at least between the extreme cases of neutrino spectrum.

In addition to the spectrum, the observation of the \df\
will allow to measure the product of the neutrino luminosity and of
$R_{SN}(0)$. This measurement will be combined with direct supernova
observations to reduce the uncertainty on $R_{SN}(0)$.  

  In the exciting case in which data from a new galactic supernova
become available before the \df\ is measured, the study of the \df\
will take a completely different direction.  The high statistics data
from the galactic supernova will give the neutrino spectrum and
luminosity with very good precision.  These and the value of $\beta$
from supernova surveys will be used as input in the analysis of the
\df\ to extract information on the less precise parameter $R_{SN}(0)$.

Depending on the precision on the \df\ signal, a number of interesting
tests could be performed.  One of them is to look for differences
between the value of $R_{SN}(0)$ estimated in astronomy and the one
obtained from neutrinos: the comparison could reveal a number of
``failed'' supernovae: objects that undergo core collapse with the
emission of neutrinos without a visible explosion.  It would also be
interesting to look for differences between the neutrino spectrum
inferred from an individual galactic supernova and the one extracted
from the \df: such differences will be a measure of the variation of
the neutrino output from one supernova to another.  Differences could
be expected in consideration of the fact that the \df\ receives the
largest contribution from low mass supernovae, $M \sim 8 -10
M_{\odot}$ (simply because the initial mass function goes roughly like
the power $ -2$ of the star's mass, see e.g. \cite{Hopkins:2006bw} and
references therein), whose neutrino output could differ from that of a
supernova with a larger mass progenitor.

In summary, the present study shows the relevance of studying the
energy spectrum of the diffuse supernova neutrino flux, mainly as a
tool to reconstruct the original neutrino spectrum and therefore to
test the physics of core collapse that influences the neutrino
spectrum formation inside the star.   It also motivates the experimental effort to reduce the background in the relevant energy window in order to achieve the necessary precision for spectral sensitivity.
\\


I  acknowledge
support from the INT-SCiDAC grant number DE-FC02-01ER41187.


\begin{thebibliography}{30}
\expandafter\ifx\csname natexlab\endcsname\relax\def\natexlab#1{#1}\fi
\expandafter\ifx\csname bibnamefont\endcsname\relax
  \def\bibnamefont#1{#1}\fi
\expandafter\ifx\csname bibfnamefont\endcsname\relax
  \def\bibfnamefont#1{#1}\fi
\expandafter\ifx\csname citenamefont\endcsname\relax
  \def\citenamefont#1{#1}\fi
\expandafter\ifx\csname url\endcsname\relax
  \def\url#1{\texttt{#1}}\fi
\expandafter\ifx\csname urlprefix\endcsname\relax\def\urlprefix{URL }\fi
\providecommand{\bibinfo}[2]{#2}
\providecommand{\eprint}[2][]{\url{#2}}

\bibitem[{\citenamefont{Arnaud et~al.}(2004)}]{Arnaud:2003zr}
\bibinfo{author}{\bibfnamefont{N.}~\bibnamefont{Arnaud}} \bibnamefont{et~al.},
  \bibinfo{journal}{Astropart. Phys.} \textbf{\bibinfo{volume}{21}},
  \bibinfo{pages}{201} (\bibinfo{year}{2004}), \eprint{gr-qc/0307101}.

\bibitem[{\citenamefont{Ando et~al.}(2005)\citenamefont{Ando, Beacom, and
  Yuksel}}]{Ando:2005ka}
\bibinfo{author}{\bibfnamefont{S.}~\bibnamefont{Ando}},
  \bibinfo{author}{\bibfnamefont{J.~F.} \bibnamefont{Beacom}},
  \bibnamefont{and} \bibinfo{author}{\bibfnamefont{H.}~\bibnamefont{Yuksel}},
  \bibinfo{journal}{Phys. Rev. Lett.} \textbf{\bibinfo{volume}{95}},
  \bibinfo{pages}{171101} (\bibinfo{year}{2005}), \eprint{astro-ph/0503321}.

\bibitem[{\citenamefont{Malek et~al.}(2003)}]{Malek:2002ns}
\bibinfo{author}{\bibfnamefont{M.}~\bibnamefont{Malek}} \bibnamefont{et~al.}
  (\bibinfo{collaboration}{Super-Kamiokande}), \bibinfo{journal}{Phys. Rev.
  Lett.} \textbf{\bibinfo{volume}{90}}, \bibinfo{pages}{061101}
  (\bibinfo{year}{2003}), \eprint{hep-ex/0209028}.

\bibitem[{\citenamefont{Eguchi et~al.}(2004)}]{Eguchi:2003gg}
\bibinfo{author}{\bibfnamefont{K.}~\bibnamefont{Eguchi}} \bibnamefont{et~al.}
  (\bibinfo{collaboration}{KamLAND}), \bibinfo{journal}{Phys. Rev. Lett.}
  \textbf{\bibinfo{volume}{92}}, \bibinfo{pages}{071301}
  (\bibinfo{year}{2004}), \eprint{hep-ex/0310047}.

\bibitem[{\citenamefont{Lunardini}(2006{\natexlab{a}})}]{Lunardini:2006sn}
\bibinfo{author}{\bibfnamefont{C.}~\bibnamefont{Lunardini}},
  \bibinfo{journal}{Phys. Rev.} \textbf{\bibinfo{volume}{D73}},
  \bibinfo{pages}{083009} (\bibinfo{year}{2006}{\natexlab{a}}),
  \eprint{hep-ph/0601054}.

\bibitem[{\citenamefont{Aharmim et~al.}(2006)}]{Aharmim:2006wq}
\bibinfo{author}{\bibfnamefont{B.}~\bibnamefont{Aharmim}} \bibnamefont{et~al.}
  (\bibinfo{collaboration}{SNO}) (\bibinfo{year}{2006}),
  \eprint{hep-ex/0607010}.

\bibitem[{\citenamefont{Beacom and Vagins}(2004)}]{Beacom:2003nk}
\bibinfo{author}{\bibfnamefont{J.~F.} \bibnamefont{Beacom}} \bibnamefont{and}
  \bibinfo{author}{\bibfnamefont{M.~R.} \bibnamefont{Vagins}},
  \bibinfo{journal}{Phys. Rev. Lett.} \textbf{\bibinfo{volume}{93}},
  \bibinfo{pages}{171101} (\bibinfo{year}{2004}), \eprint{hep-ph/0309300}.

\bibitem[{\citenamefont{Yuksel et~al.}(2006)\citenamefont{Yuksel, Ando, and
  Beacom}}]{Yuksel:2005ae}
\bibinfo{author}{\bibfnamefont{H.}~\bibnamefont{Yuksel}},
  \bibinfo{author}{\bibfnamefont{S.}~\bibnamefont{Ando}}, \bibnamefont{and}
  \bibinfo{author}{\bibfnamefont{J.~F.} \bibnamefont{Beacom}},
  \bibinfo{journal}{Phys. Rev.} \textbf{\bibinfo{volume}{C74}},
  \bibinfo{pages}{015803} (\bibinfo{year}{2006}), \eprint{astro-ph/0509297}.

\bibitem[{\citenamefont{Nakamura}(2003)}]{Nakamura:2003hk}
\bibinfo{author}{\bibfnamefont{K.}~\bibnamefont{Nakamura}},
  \bibinfo{journal}{Int. J. Mod. Phys.} \textbf{\bibinfo{volume}{A18}},
  \bibinfo{pages}{4053} (\bibinfo{year}{2003}).

\bibitem[{\citenamefont{Jung}(2000)}]{Jung:1999jq}
\bibinfo{author}{\bibfnamefont{C.~K.} \bibnamefont{Jung}},
  \bibinfo{journal}{AIP Conf. Proc.} \textbf{\bibinfo{volume}{533}},
  \bibinfo{pages}{29} (\bibinfo{year}{2000}), \eprint{hep-ex/0005046}.

\bibitem[{\citenamefont{Mosca}(2005)}]{Mosca:2005mi}
\bibinfo{author}{\bibfnamefont{L.}~\bibnamefont{Mosca}},
  \bibinfo{journal}{Nucl. Phys. Proc. Suppl.} \textbf{\bibinfo{volume}{138}},
  \bibinfo{pages}{203} (\bibinfo{year}{2005}).

\bibitem[{\citenamefont{Marrodan~Undagoitia
  et~al.}(2006)}]{MarrodanUndagoitia:2006re}
\bibinfo{author}{\bibfnamefont{T.}~\bibnamefont{Marrodan~Undagoitia}}
  \bibnamefont{et~al.}, \bibinfo{journal}{Prog. Part. Nucl. Phys.}
  \textbf{\bibinfo{volume}{57}}, \bibinfo{pages}{283} (\bibinfo{year}{2006}),
  \eprint{hep-ph/0605229}.


\bibitem[{\citenamefont{Wurm
  et~al.}(2007)}]{Wurm:2007cy}
\bibinfo{author}{\bibfnamefont{M.}~\bibnamefont{Wurm}}
  \bibnamefont{et~al.}, \bibinfo{journal}{Phys. Rev. D}
  \textbf{\bibinfo{volume}{75}}, \bibinfo{pages}{023007} (\bibinfo{year}{2007}),
  \eprint{astro-ph/0701305}.



\bibitem[{\citenamefont{Ereditato and Rubbia}(2006)}]{Ereditato:2005yx}
\bibinfo{author}{\bibfnamefont{A.}~\bibnamefont{Ereditato}} \bibnamefont{and}
  \bibinfo{author}{\bibfnamefont{A.}~\bibnamefont{Rubbia}},
  \bibinfo{journal}{Nucl. Phys. Proc. Suppl.} \textbf{\bibinfo{volume}{155}},
  \bibinfo{pages}{233} (\bibinfo{year}{2006}), \eprint{hep-ph/0510131}.

\bibitem[{\citenamefont{Cline et~al.}(2006)\citenamefont{Cline, Raffaelli, and
  Sergiampietri}}]{Cline:2006st}
\bibinfo{author}{\bibfnamefont{D.~B.} \bibnamefont{Cline}},
  \bibinfo{author}{\bibfnamefont{F.}~\bibnamefont{Raffaelli}},
  \bibnamefont{and}
  \bibinfo{author}{\bibfnamefont{F.}~\bibnamefont{Sergiampietri}},
  \bibinfo{journal}{JINST} \textbf{\bibinfo{volume}{1}},
  \bibinfo{pages}{T09001} (\bibinfo{year}{2006}), \eprint{astro-ph/0604548}.

\bibitem[{\citenamefont{Fogli et~al.}(2005)\citenamefont{Fogli, Lisi, Mirizzi,
  and Montanino}}]{Fogli:2004ff}
\bibinfo{author}{\bibfnamefont{G.~L.} \bibnamefont{Fogli}},
  \bibinfo{author}{\bibfnamefont{E.}~\bibnamefont{Lisi}},
  \bibinfo{author}{\bibfnamefont{A.}~\bibnamefont{Mirizzi}}, \bibnamefont{and}
  \bibinfo{author}{\bibfnamefont{D.}~\bibnamefont{Montanino}},
  \bibinfo{journal}{JCAP} \textbf{\bibinfo{volume}{0504}}, \bibinfo{pages}{002}
  (\bibinfo{year}{2005}), \eprint{hep-ph/0412046}.

\bibitem[{\citenamefont{Dighe and Smirnov}(2000)}]{Dighe:1999bi}
\bibinfo{author}{\bibfnamefont{A.~S.} \bibnamefont{Dighe}} \bibnamefont{and}
  \bibinfo{author}{\bibfnamefont{A.~Y.} \bibnamefont{Smirnov}},
  \bibinfo{journal}{Phys. Rev.} \textbf{\bibinfo{volume}{D62}},
  \bibinfo{pages}{033007} (\bibinfo{year}{2000}), \eprint{hep-ph/9907423}.

\bibitem[{\citenamefont{Lunardini and Smirnov}(2003)}]{Lunardini:2003eh}
\bibinfo{author}{\bibfnamefont{C.}~\bibnamefont{Lunardini}} \bibnamefont{and}
  \bibinfo{author}{\bibfnamefont{A.~Y.} \bibnamefont{Smirnov}},
  \bibinfo{journal}{JCAP} \textbf{\bibinfo{volume}{0306}}, \bibinfo{pages}{009}
  (\bibinfo{year}{2003}), \eprint{hep-ph/0302033}.

\bibitem[{\citenamefont{Keil et~al.}(2003)\citenamefont{Keil, Raffelt, and
  Janka}}]{Keil:2002in}
\bibinfo{author}{\bibfnamefont{M.~T.} \bibnamefont{Keil}},
  \bibinfo{author}{\bibfnamefont{G.~G.} \bibnamefont{Raffelt}},
  \bibnamefont{and} \bibinfo{author}{\bibfnamefont{H.-T.} \bibnamefont{Janka}},
  \bibinfo{journal}{Astrophys. J.} \textbf{\bibinfo{volume}{590}},
  \bibinfo{pages}{971} (\bibinfo{year}{2003}), \eprint{astro-ph/0208035}.

\bibitem[{\citenamefont{Lunardini}(2006{\natexlab{b}})}]{Lunardini:2005jf}
\bibinfo{author}{\bibfnamefont{C.}~\bibnamefont{Lunardini}},
  \bibinfo{journal}{Astropart. Phys.} \textbf{\bibinfo{volume}{26}},
  \bibinfo{pages}{190} (\bibinfo{year}{2006}{\natexlab{b}}),
  \eprint{astro-ph/0509233}.

\bibitem[{\citenamefont{Hopkins and Beacom}(2006)}]{Hopkins:2006bw}
\bibinfo{author}{\bibfnamefont{A.~M.} \bibnamefont{Hopkins}} \bibnamefont{and}
  \bibinfo{author}{\bibfnamefont{J.~F.} \bibnamefont{Beacom}},
  \bibinfo{journal}{Astrophys. J.} \textbf{\bibinfo{volume}{651}},
  \bibinfo{pages}{142} (\bibinfo{year}{2006}), \eprint{astro-ph/0601463}.

\bibitem[{\citenamefont{Hernquist and Springel}(2003)}]{Hernquist:2002rg}
\bibinfo{author}{\bibfnamefont{L.}~\bibnamefont{Hernquist}} \bibnamefont{and}
  \bibinfo{author}{\bibfnamefont{V.}~\bibnamefont{Springel}},
  \bibinfo{journal}{Mon. Not. Roy. Astron. Soc.}
  \textbf{\bibinfo{volume}{341}}, \bibinfo{pages}{1253} (\bibinfo{year}{2003}),
  \eprint{astro-ph/0209183}.

\bibitem[{\citenamefont{Ando and Sato}(2004)}]{Ando:2004hc}
\bibinfo{author}{\bibfnamefont{S.}~\bibnamefont{Ando}} \bibnamefont{and}
  \bibinfo{author}{\bibfnamefont{K.}~\bibnamefont{Sato}}, \bibinfo{journal}{New
  J. Phys.} \textbf{\bibinfo{volume}{6}}, \bibinfo{pages}{170}
  (\bibinfo{year}{2004}), \eprint{astro-ph/0410061}.

\bibitem[{\citenamefont{Malek}(2003)}]{Malek:2003ki}
\bibinfo{author}{\bibfnamefont{M.~S.} \bibnamefont{Malek}}
  (\bibinfo{year}{2003}), \bibinfo{note}{uMI-31-06530, PhD thesis, available at
  http://www-sk.icrr.u-tokyo.ac.jp/sk/index-e.html}.

\bibitem[{\citenamefont{Strumia and Vissani}(2003)}]{Strumia:2003zx}
\bibinfo{author}{\bibfnamefont{A.}~\bibnamefont{Strumia}} \bibnamefont{and}
  \bibinfo{author}{\bibfnamefont{F.}~\bibnamefont{Vissani}},
  \bibinfo{journal}{Phys. Lett.} \textbf{\bibinfo{volume}{B564}},
  \bibinfo{pages}{42} (\bibinfo{year}{2003}), \eprint{astro-ph/0302055}.

\bibitem[{\citenamefont{Hirata et~al.}(1988)}]{Hirata:1988ad}
\bibinfo{author}{\bibfnamefont{K.~S.} \bibnamefont{Hirata}}
  \bibnamefont{et~al.}, \bibinfo{journal}{Phys. Rev.}
  \textbf{\bibinfo{volume}{D38}}, \bibinfo{pages}{448} (\bibinfo{year}{1988}).

\bibitem[{\citenamefont{Bahcall et~al.}(1997)\citenamefont{Bahcall, Krastev,
  and Lisi}}]{Bahcall:1996ha}
\bibinfo{author}{\bibfnamefont{J.~N.} \bibnamefont{Bahcall}},
  \bibinfo{author}{\bibfnamefont{P.~I.} \bibnamefont{Krastev}},
  \bibnamefont{and} \bibinfo{author}{\bibfnamefont{E.}~\bibnamefont{Lisi}},
  \bibinfo{journal}{Phys. Rev.} \textbf{\bibinfo{volume}{C55}},
  \bibinfo{pages}{494} (\bibinfo{year}{1997}), \eprint{nucl-ex/9610010}.

\bibitem[{\citenamefont{JWST}()}]{jwst}
\bibinfo{author}{\bibnamefont{JWST}} \bibinfo{note}{web page,
  http://www.jwst.nasa.gov}.

\bibitem[{\citenamefont{SNAP}()}]{snap}
\bibinfo{author}{\bibnamefont{SNAP}} \bibinfo{note}{letter of intent,
  available at http://snap.lbl.gov}.

\bibitem[{\citenamefont{Liebendoerfer et~al.}(2003)}]{Liebendoerfer:2002ny}
\bibinfo{author}{\bibfnamefont{M.}~\bibnamefont{Liebendoerfer}}
  \bibnamefont{et~al.}, \bibinfo{journal}{Nucl. Phys.}
  \textbf{\bibinfo{volume}{A719}}, \bibinfo{pages}{144} (\bibinfo{year}{2003}),
  \eprint{astro-ph/0211329}.

\bibitem[{\citenamefont{Kitaura et~al.}(2005)\citenamefont{Kitaura, Janka, and
  Hillebrandt}}]{Kitaura:2005bt}
\bibinfo{author}{\bibfnamefont{F.~S.} \bibnamefont{Kitaura}},
  \bibinfo{author}{\bibfnamefont{H.-T.} \bibnamefont{Janka}}, \bibnamefont{and}
  \bibinfo{author}{\bibfnamefont{W.}~\bibnamefont{Hillebrandt}}
  (\bibinfo{year}{2005}), \eprint{astro-ph/0512065}.

\end{thebibliography}

\end{document}